\documentclass[11pt]{article}

\usepackage{graphicx}

\newlength{\bredde}
\def\slash#1{\settowidth{\bredde}{$#1$}\ifmmode\,\raisebox{.15ex}{/}
\hspace*{-\bredde} #1\else$\,\raisebox{.15ex}{/}\hspace*{-\bredde} #1$\fi}
\newcommand{\beq}{\begin{equation}}
\newcommand{\eeq}{\end{equation}}
\newcommand{\ba}{\begin{array}{ccc}}
\newcommand{\ea}{\end{array}}

\newcommand{\noi}{\vspace{12pt}\noindent}
\newcommand{\lG}{\raise.3ex\hbox{$\stackrel{\leftarrow}{G}$}}
\newcommand{\lU}{\raise.3ex\hbox{$\stackrel{\leftarrow}{U}$}}
\newcommand{\lP}{\raise.3ex\hbox{$\stackrel{\leftarrow}{{\cal P}}$}}
\newcommand{\leta}{\raise.3ex\hbox{$\stackrel{\leftarrow}{\eta}$}}
\newcommand{\lOmega}{\raise.3ex\hbox{$\stackrel{\leftarrow}{\Omega}$}}
\newcommand{\ldr}{\raise.3ex\hbox{$\stackrel{\leftarrow}{\delta^r}$}}

\def\m2{{\mathcal{M}}^{\dagger}{\mathcal{M}}}
\def\mb2{M^2}

\newcommand{\unit}{1\!\!\!\!1}
\def\beqn{\begin{eqnarray}}
\def\eeqn{\end{eqnarray}}

\def\gtwid{\raise.3ex\hbox{$>$\kern-.75em\lower1ex\hbox{$\sim$}}}
\def\ltwid{\raise.3ex\hbox{$<$\kern-.75em\lower1ex\hbox{$\sim$}}}

\begin{document}
\topmargin -1.4cm
\oddsidemargin -0.8cm
\evensidemargin -0.8cm

\title{
{\vspace{-1.5cm} \normalsize
\hfill \parbox{40mm}{CERN-TH/2001-268}}\\
{\vspace{-0.35cm} \normalsize
\hfill \parbox{40mm}{FSU-CSIT-01-47}}\\
{\vspace{-0.35cm} \normalsize
\hfill \parbox{40mm}{COLO-HEP-473}}\\
{\vspace{-0.35cm} \normalsize
\hfill \parbox{40mm}{MIT-CTP 3196}}\\[5mm]
\Large{{\bf Patterns of Spontaneous Chiral Symmetry Breaking\\
in Vectorlike Gauge Theories}}}

\vspace{1.5cm}

\author{{\sc P. H. Damgaard}\footnote{On leave from: the Niels Bohr institute,
Blegdamsvej 17, DK-2100 Copenhagen, Denmark.}
\\TH-Division, CERN\\CH-1211 Geneva 23,
Switzerland\\~\\{\sc U.M. Heller}\\CSIT, Florida State University\\
Tallahassee FL 32306-4120, U.S.A.\\~\\{\sc R. Niclasen}\\Physics Department, 
University of Colorado\\Boulder CO 80309, U.S.A.\\~\\{\sc B. Svetitsky}\\
Center for Theoretical Physics\\ 
Laboratory of Nuclear Science and Department of Physics\\
Massachusetts Institute of Technology\\Cambridge, MA 02139,
USA\\and\\School of Physics and Astronomy\\
Raymond and Beverly Sackler Faculty
  of Exact Sciences\\
Tel Aviv University\\
69978 Tel Aviv, Israel\footnote{Permanent address.}}
\date{\today}
\maketitle
\vfill
\begin{abstract}
It has been conjectured that spontaneous chiral symmetry breaking in
strongly coupled vectorlike gauge theories falls into only three different
classes, depending on the gauge group and the representations carried
by the fermions. We test this proposal by studying SU(2), SU(3) and
SU(4) lattice gauge theories with staggered fermions in different
irreducible representations. Staggered fermions away from the continuum
limit should, for all complex representations, still belong to the continuum
class of spontaneous symmetry breaking. But for all real and pseudo-real
representations we show that staggered fermions should belong to incorrect
symmetry breaking classes away from the continuum, thus generalizing
previous results. As an unambiguous signal for whether chiral
symmetry breaks, and which breaking pattern it follows, we look at the
smallest Dirac eigenvalue distributions. We find that the patterns of
symmetry breaking are precisely those conjectured.
\end{abstract}
\vfill

\thispagestyle{empty}
\newpage

\section{Introduction}

\noi
Consider an asymptotically free gauge theory of gauge group ${\cal G}$
coupled vectorially to $N_f$ massless Dirac fermions that all transform
according to some irreducible representation $r$ of the gauge
group. This theory will have a global chiral-flavor symmetry, whose
group we denote by $G$. An obvious general question to ask is whether
this group $G$ will be spontaneously broken down to some subgroup $H$,
and if so, what is the precise symmetry breaking pattern.
This question is relevant for the construction of technicolor theories,
but it is also of interest in its own right. Conventional folklore says
that there are, for four-dimensional field theories, just three
classes of breaking to consider \cite{Peskin}:
\begin{itemize}
\item The fermion representation $r$ is pseudo-real: Chiral symmetries
are enhanced from $SU(N_f)\times SU(N_f)$ to $SU(2N_f)$, and the
expected
symmetry breaking pattern is $SU(2N_f) \to Sp(2N_f)$.
\item The fermion representation $r$ is complex: The expected
symmetry breaking
pattern is $SU(N_f)\times SU(N_f) \to SU(N_f)$.
\item The fermion representation $r$ is real: Chiral symmetries are again
enhanced to $SU(2N_f)$, and the expected symmetry breaking pattern is
$SU(2N_f) \to SO(2N_f)$.
\end{itemize}
This remarkably simple classification of the possible symmetry breaking
patterns is what follows from the assumption of maximal
breaking of chiral symmetry with simultaneous preservation of
maximal flavor symmetry. Indeed, spontaneous flavor symmetry breaking
in vectorlike gauge theories is prohibited by the
Vafa-Witten theorem~\cite{VW}, and the above symmetry breaking patterns
are thus consistent with this theorem. Very recently it has been
checked that the most promising candidate for a $c$-theorem in
four dimensions is consistent with these breaking schemes for all
simple gauge groups ${\cal G}$ and fermions transforming according
to arbitrary irreducible representations under those gauge groups
\cite{BD}. At a large number of colors, the Coleman-Witten argument
\cite{CW} and a simple generalization of it to pseudo-real gauge
groups \cite{BD} shows that indeed the symmetry breaking patterns
are precisely those expected. But a general proof for a finite number
of colors is still lacking.

\noi
One way to test whether the dynamics really is such these three types of
symmetry breaking do occur is to compute the appropriate chiral
condensates by numerical means in lattice gauge theory. This is
unfortunately a highly non-trivial task. Consider the definition
of a chiral condensate
\beq
\Sigma ~=~ \lim_{m\to 0}\lim_{V\to 0}\langle \bar{\Psi}\Psi\rangle
\label{sigmadef}
\eeq
which requires the limit of infinite four-volume $V\to\infty$ to be
taken prior to the massless limit $m\to 0$. While it is already
increasingly time-consuming to simulate lighter and lighter fermions
by Monte Carlo methods, the infinite volume limit obviously cannot
be taken under any circumstances. Even in the quenched case, where
the massless limit can be taken trivially, the above order parameter
for spontaneous chiral symmetry breaking can only be used in
conjunction with an often quite uncertain extrapolation to the infinite
volume limit.

\noi
Instead, we shall here make use of a much more efficient technique. It
is based on the fact that the distributions of the smallest Dirac
operator eigenvalues, when measured over different volumes, contain
the combined information of 1) whether spontaneous chiral symmetry
breaking occurs, and 2) which symmetry breaking pattern is realized.
So far, this has been confirmed for the fundamental and adjoint
representations of SU(2) and SU(3) gauge groups \cite{BBetal}.
Tests have also been
performed on strongly coupled U(1) theories, in both 2 and 4 dimensions
\cite{Lang}. Here, we shall be
a little more systematic, and probe also more exotic irreducible
representations. If the above conjecture about spontaneous symmetry breaking
is correct, we know precisely which patterns the different cases should
follow. Because the distributions of the smallest Dirac operator
eigenvalues are exact finite-volume scaling functions, this turns the
finite volumes used in lattice gauge theory simulations into a
distinct advantage.

\noi
We shall here briefly review how the distribution of the smallest
Dirac operator eigenvalues can be used to show both that spontaneous
chiral symmetry breaking occurs {\em and} specify uniquely to which symmetry
breaking class it belongs. In fact, it is obvious that we need only
consider the smallest Dirac operator eigenvalues because by the
Banks-Casher relation
\beq
\Sigma ~=~ \pi \rho(0)
\eeq
it is precisely the density of Dirac operator eigenvalues
around the origin which determines whether or not chiral symmetry
is spontaneously broken. At finite volume $V$ this means that the
eigenvalue spacing near the origin must go like $1/(V\rho(0))$ \cite{LS}.
This leads to the definition of the microsopically rescaled spectral
density \cite{SV},
\beq
\rho_s(\zeta) ~\equiv~ \frac{1}{V} \rho\left(\frac{\zeta}{V\Sigma}\right) ~,
\label{rhodef}
\eeq
which should have a finite limit as $V \to \infty$. Effectively, one
is blowing up the very smallest Dirac operator eigenvalues so that
the average eigenvalue spacing is of order unity, rather than
$1/(V\rho(0))$. This is finite-volume scaling: the eigenvalue
density should fall on one universal curve if the eigenvalues $\lambda$ are
rescaled according to $\zeta \equiv V\Sigma \lambda$.
So to demonstrate spontaneous chiral symmetry breaking it suffices
to check that the smallest Dirac operator eigenvalues indeed do
satisfy this simple finite-volume scaling, even if the functional
form of (\ref{rhodef}) were not known.
Actually, the microscopic spectral densities and higher spectral
correlation functions {\em are} in fact
known analytically from universal Random Matrix Theory results (see
\cite{VZ,ADMN,Zetal,DN,WGW} for the $\beta = 2$ universality class,
and \cite{V,SenerV,AD,NN,AK,FG} for the two other universality classes).
This holds in arbitrary sectors of fixed topological charge $\nu$.
It is
now also known how these results can be derived directly from field theory
\cite{OTV}, although the method based on Random Matrix Theory remains
the simplest.

\noi
The same argument holds for the distribution of individual Dirac operator
eigenvalues, again rescaled to microscopic variables $\zeta =
V\Sigma \lambda$. Analytical formulas exist for an infinite sequence of such
eigenvalue distributions for all three universality classes \cite{NDW}.
The only exception is the $\beta=1$
case, where so far only the case of odd topological charge $\nu$
has been computed in all generality, but here at least the lowest
eigenvalue distribution is known for the most interesting case, namely
$\nu=0$ \cite{F}. As both the spectral correlation functions, and
the individual eigenvalue distributions differ markedly for the three
different universality classes, all of these observables are excellent
candidates for determining the chiral symmetry breaking pattern. The
analytical functions with which to compare are generally rather simple
combinations of Bessel functions.

\section{Symmetries of the Dirac Operator and Chiral Symmetry Breaking}

\noi
To make the connection between the three symmetry breaking classes and the
Random Matrix Theory results, it is instructive to first work out the
connection between (anti-unitary)
symmetries of the Dirac operator, and the type of
representation the fermions carry ($i.e.$, whether they are pseudo-real,
complex, or real).
In fact, this argument is only to make the connection to Random Matrix
Theory. We know already from the chiral Lagrangian alone that the 
microscopic Dirac operator spectrum
always matches one-to-one the three symmetry breaking classes above.
If Random Matrix Theory is to give equivalent results as those that
can be derived from field theory, $i.e.$, from the relation between
the fundamental Lagrangians and those of the corresponding
effective chiral Lagrangians, it has to be determined by just these three
symmetry breaking classes.
What we learn from this is that there must exist a direct relation
between the anti-unitary symmetries of the Dirac operator and the
pattern of chiral symmetry breaking, in all generality. This is not too
surprising, since the pattern of symmetry breaking is just given by
the color representation of the fermions, and we shall return to this
connection below.

\noi
The relevant question is whether there is an anti-unitary
symmetry $V K$ of the Dirac operator,
\beq
              [V K, D(A)] ~=~ 0 ~,
\eeq
where $K$ is the operation of complex conjugation and $V$
is unitary.
The generic case is that there is no such symmetry, and then the
matrix elements of the Dirac operator in a chiral basis are simply complex.
The corresponding Random Matrix Theory ensemble is (chiral)
unitary, {\it ie.,} $\beta=2$ \cite{Vthreefold,SmV}.

\noi
If such a symmetry exists, then either $(V K)^2 = +1$ or $(V K)^2 = -1$,
and both $\psi$ and $V K \psi$ have the same eigenvalues of $D(A)$.
When $(V K)^2 = +1$ the two eigenvectors $\psi$ and $V K\psi$
are, however, not linearly independent ($V K \psi = \pm \psi$). But if
$(V K)^2 = -1$, the two eigenvectors {\em are} linearly independent, and
we have a genuine double degeneracy of eigenvalues. In the former case
a basis exists so that the matrix elements of the Dirac operator are
real, and in Random Matrix Theory language this
corresponds to the (chiral) orthogonal ($\beta=1$) ensemble \cite{Vthreefold}.
In the latter case the matrix elements of the Dirac operator can be
diagonalized by a symplectic transformation, and the corresponding
Random Matrix Theory is the (chiral) symplectic ($\beta=4$) one
\cite{Vthreefold}.

\noi
Now, the complex conjugation $K$ affects both the gauge fields and the
gamma matrices in the Dirac operator.
Let $\Gamma$ be the $\gamma$-matrix for which
$\Gamma \gamma_\mu^* \Gamma^\dagger
= \gamma_\mu$. Then $\Gamma$ is the part of $V$, if $V$ exists, which
acts on the Dirac indices. For both the Euclidean Dirac and chiral
representation of the $\gamma$-matrices we have $\Gamma = \gamma_1 \gamma_3$.
So we write $V = \Gamma S$, where $S$ acts entirely in color space.
Then $(V K)^2 = V K V K = V V^* K K = V V^*$, since $K^2 = 1$, and with
$V = \Gamma S$ we have $(V K)^2 = S S^* \Gamma \Gamma^*$.
For $\Gamma = \gamma_1 \gamma_3$, $\Gamma^* = \Gamma$ and $\Gamma^2 = -1$.
Therefore we find $(V K)^2 = - S S^*$.

\noi
For the gauge field part, let us remind ourselves of some basic properties
of complex, real and pseudo-real representations. A {\em complex}
representation is one which is not equivalent to its complex conjugate.
A {\em real} representation is equivalent to its complex conjugate in the
following sense. Let $T^a$ be the (hermitian) generators in the given
representation. It is real if there exists a unitary $S$ so that
\beq
\left(T^a\right)^* ~=~ \left(T^a\right)^T ~=~ - S^{-1}T^aS ~~~,~~~~~~~
 SS^* ~=~ 1
\label{real}
\eeq
For such a representation, $U = \exp[i\omega^aT^a]$ is equivalent to its
complex conjugate (equal after a similarity transformation).
It is then possible to find a basis in which all $T^a$ are purely imaginary
(but of course still Hermitian). A {\em pseudo-real} representation is one
for which there exists a unitary $S$ so that
\beq
\left(T^a\right)^* ~=~ \left(T^a\right)^T ~=~ - S^{-1}T^aS ~~~,~~~~~~~
 SS^* ~=~ -1
\label{pseudo}
\eeq
It is then no longer possible to find a basis in which all generators
$T^a$ are purely imaginary. The fundamental representation of
SU(2) with its conventional Pauli matrices $\tau^a/2$ is a good example.
Since (\ref{real}) implies $S = S^T$, and (\ref{pseudo}) implies $S=-S^T$,
the one major distinction between real and pseudo-real is whether $S$ is
{\em symmetric} or {\em antisymmetric}.

\noi
For a {\em real} representation we thus find $(V K)^2 = -1$ and the Dirac
operator will belong to the {\em symplectic} ensemble, $\beta=4$, while
for a {\em pseudo-real} representation $(V K)^2 = 1$ and the Dirac
operator belongs to the {\em orthogonal} ensemble, $\beta=1$. This is a
little counter-intuitive, compared to the symmetry breaking patterns. The
reason is the behavior of the $\gamma$-matrices in the Dirac operator under
complex conjugation.

\noi
So far we have only been concerned with the symmetries of the (massless)
Dirac operator. In fact there is an intimate connection between the
symmetries of the Dirac operator and the pattern of spontaneous symmetry
breaking. This may seem surprising, since the conjectured symmetry
breaking patterns are based on the symmetries of the condensate
$\langle\bar{\Psi}\Psi\rangle$ rather than those of the Dirac operator.
In order to elucidate the relation between the two, let us first quickly
recall how the appearance of the three different patterns of spontaneous
chiral symmetry breaking can be understood \cite{Peskin}. For this
purpose it is convenient to introduce the two-component van der Waerden
notation of dotted and undotted spinor indices, so familiar from $e.g.$
supersymmetry. We consider 4-component massless Dirac spinors
$\Psi(x)$, and choose to work in a chiral basis of $\gamma$-matrices.
Then upper and lower parts of the 4-spinors simply correspond to the
left-handed and right-handed components:
\beq
\Psi ~=~ \pmatrix{ \Psi_L\cr \Psi_R } ~\equiv~ \pmatrix{ \psi_{\alpha}
\cr \bar{\chi}^{\dot{\beta}} }
~=~ \pmatrix{ \psi_{\alpha}
\cr (\chi^{\beta})^*} ~, \label{lrdef}
\eeq
and the charge conjugate spinor is then given by
\beq
\Psi^C ~=~  C\bar{\Psi}^T ~=~ -i\gamma^0\gamma^2\bar{\Psi}^T
~=~ \pmatrix{ \chi_{\alpha}\cr \bar{\psi}^{\dot{\beta}}} ~.
\eeq
In other words, instead of working with one 4-component Dirac spinor
$\Psi$, we can equally well work with two left-handed spinors from $\Psi$
and its charge conjugate $\Psi^C$,
\beq
\psi_{\alpha}~~~{\mbox{\rm and}}~~~ \chi_{\beta} ~=~ \epsilon_{\beta\gamma}
\chi^{\gamma} ~,
\eeq
where in the last equation we have made use of the fact that the two-component
spinor indices are raised and lowered by means of the antisymmetric
$\epsilon$-tensor.
Consider now a well-known example of spontaneous chiral symmetry breaking,
that of a QCD-like theory with $N_f$ flavors of massless quarks
transforming according to the fundamental representation of gauge
group ${\cal G}$ = SU(3). An order parameter is the well-known condensate
of $\bar{\Psi}\Psi$, which includes a summation of both color and flavor
indices. Let us make this explicit, and at the same time write the fermion
bilinear in terms of the two-component spinors:
\beq
\bar{\Psi}\Psi ~=~ \epsilon^{\alpha\beta}\chi^{ia}_{\beta}\psi_{\alpha ia}
~~ + ~~{\mbox{\rm h.c.}}
\eeq
Here $\alpha$ and $\beta$ are the two-component spinor indices, while
$i$ and $a$ denote flavor and color indices, respectively. Now,
$\psi_{\alpha ia}$ transforms like a 3-representation under color
while $\chi^{\beta ia}$ transforms like a $\bar{3}$ (see eq. (\ref{lrdef})).
It is thus convenient to relabel the latter spinor as a $\psi$-spinor
with the color transformation property made explicit:
\beq
\bar{\Psi}\Psi ~=~ \epsilon^{\alpha\beta}\psi^{i(\bar{3})}_{\beta}
\psi^{(3)}_{\alpha i} ~~
+ ~~{\mbox{\rm h.c.}}
\eeq
Then it is immediately clear that the generalization to an arbitrary
complex representation $r$ of gauge group ${\cal G}$ is
$$
\epsilon^{\alpha\beta}\psi^{i(\bar{r})}_{\beta}
\psi^{(r)}_{\alpha i} ~~
+ ~~{\mbox{\rm h.c.}}
$$
Note how the left-handed and right-handed pieces trivially are invariant
under the same symmetries, since the right-handed part is just the
Hermitian conjugate of the left-handed part.
The term above is in fact the ${\cal G}$-invariant fermion bilinear of maximum
vectorlike flavor symmetry, and if it attains a non-vanishing expectation
value it is thus consistent with the Vafa-Witten theorem. The flavor symmetry
remaining of the above expression is only $SU(N_f)$, and the symmetry
breaking pattern, if realized, thus corresponds to
\beq
SU(N_f)\times SU(N_f) ~\to~ SU(N_f)
\eeq
for all complex representations.

\noi
For real representations $r$ of the gauge group ${\cal G}$ the representation
$r$ is equivalent to its complex conjugate $\bar{r}$.
The initial symmetry is then bigger,
enlarged to $SU(2N_f)$ because $\psi$ and $S\chi$ (with $S$
symmetric, as discussed above) transform
in the same way under color, and thus can mix. The ${\cal G}$-invariant
fermion bilinear of maximal flavor symmetry is then
$$
\epsilon^{\alpha\beta}\psi^{ia}_{\beta}
\psi^{b}_{\alpha i}S^{-1}_{ab}
$$
where $S$ is the symmetric matrix described above. Because of fermi
statistics this bilinear can have non-vanishing expectation value. The
continuous flavor symmetries remaining are only those of orthogonal
transformations, so the symmetry breaking in that case should be
\beq
SU(2N_f) ~\to~ SO(2N_f)
\eeq
Because of the doubling in symmetries it is in this case possible to
consider also the breaking pattern of Majorana fermions, which effectively
corresponds to replacing $2N_f$ by $N_f$ (real, Majorana fermions).

\noi
Finally there is the pseudo-real case. Although the representation
$r$ in that case is not equivalent to $\bar{r}$, it is possible to arrange
for a fermion transforming according to, say, $\bar{r}$ to transform
according to $r$ by multiplying by the antisymmetric matrix $S$
of eq. (\ref{pseudo}). We can thus again, by this relabelling, work with
fields that only transform according to the representation $r$.
Because of its antisymmetry, the only way to form
a non-vanishing bilinear out of anticommuting fermion fields
is by multiplying with
a matrix antisymmetric in flavor indices. The result is
$$
\epsilon^{\alpha\beta}\psi^{ia}_{\beta}
\psi^{jb}_{\alpha}S^{-1}_{ab}E_{ij}
$$
where now $S=-S^T$, and also $E=-E^T$. The group of continuous flavor
transformations
leaving this quadratic form invariant is $Sp(2N_f)$, and the expected
symmetry breaking pattern is thus
\beq
SU(2N_f) ~\to~ Sp(2N_f) ~.
\eeq

\noi
All of this is standard. What is perhaps puzzling is that these
considerations in no way involve the symmetries of the Dirac operator,
the starting point for the analysis in terms of Random Matrix Theory.
The conjectured symmetry breaking patterns were originally based on
the intuitive idea of maximally breaking chiral symmetry without breaking
flavor symmetries, an idea which subsequently found its justification in
the Vafa-Witten theorem. This suggests that the Random Matrix Theory approach,
and its associated three chiral matrix ensembles \cite{Vthreefold}, in
some way should contain the same ingredients that enter in the proof
of the Vafa-Witten theorem. This idea is not totally far-fetched
because in fact the main assumption on which the Vafa-Witten
theorem rests \cite{VW} is positivity of the measure, which for the
fermionic part can be traced back to the fact that Dirac eigenvalue
density is even in $\lambda: ~\rho(\lambda) = \rho(-\lambda)$. This
property is automatically built into the chiral Random Matrix
Theory, with $\rho(\lambda)$ now being replaced by the eigenvalue
density of the random matrices.
As for the precise symmetry breaking patterns, we have
seen that the classification in terms of Random Matrix Theory goes
parallel with the classification based on the assumption of maximal
chiral symmetry breaking (without breaking flavor symmetries) in that
it depends on the color representation only. Without
any reference to Random Matrix Theory, in a chiral basis the Dirac operator
matrix elements are complex for complex representations, can be chosen real
for pseudo-real representations, and can be chosen quaternion-real
for real representations \cite{Vthreefold}. In the latter case the
Dirac operator eigenvalues are doubly degenerate. In this sense
the classification according to the Dyson indices $\beta$ can be done
independently of the specific chiral Random Matrix Theories. The fact
that the chiral Random Matrix Theories in the microscopic limit can
be mapped to precisely the zero-momentum mode effective chiral
Largrangian corresponding to just the right cosets of symmetry
breaking \cite{HV} is a remarkable fact for which there is clearly
no simple explanation based only on group theoretic arguments.

\section{Staggered Fermions}

\noi
For staggered fermions the situation is both simpler and more complicated.
More complicated is the pattern of symmetry breaking.
Simpler are the symmetries of the Dirac operator. Since the staggered
Dirac operator does not have any $\gamma$-matrices, but only sign factors,
the (real) Kogut-Susskind phases, $\eta_\mu(x) = \pm 1$, the potential
anti-unitary symmetry operators are just $S K$, with $S$ as before acting
in color space,
\beq
              [S K, D(U)] ~=~ 0 ~.
\eeq
Now $(S K)^2 = S S^*$, and hence {\em real} representation staggered
fermions should be associated with 
the chiral {\em orthogonal} ensemble, $\beta=1$, and {\em pseudo-real}
representation staggered fermions should correspondingly be associated with
the chiral {\em symplectic} ensemble. As compared to
the continuum case the two classes $\beta=1$ and $\beta=4$ are always swapped.

\noi
The case of complex representations, when no anti-unitary symmetry operator
exists, is the same for continuum and for staggered fermions. The
corresponding Random Matrix Theory ensemble is chiral {\em unitary}, $\beta=2$.

\noi
Staggered fermions in the fundamental representation of
SU(2) are an example of the $\beta=4$, symplectic case. The anti-unitary
transformation is then given by $S K = i \sigma_2 K$, since for elements of
the gauge group SU(2) $(i \sigma_2) U^* (-i \sigma_2) = U$ and since for
staggered fermions $D(U)^* = D(U^*)$. Thus $[i\sigma_2 K,D(U)]=0$,
and, since $i\sigma_2$ is real,
$(i\sigma_2 K)^2 = (i\sigma_2)^2 = -1$, so we have the symplectic case.
We see that it crucially hinges on the relation $D(U)^* = D(U^*)$.

\noi
Another example of staggered fermions in the symplectic ensemble are
staggered fermions in the $j=3/2$ representation of SU(2).
The analogous transformation is then given by
\beq
U_{(3/2)} = \pmatrix{ 0 & 0 & 0 & -1 \cr 0 & 0 & 1 & 0 \cr
                      0 & -1 & 0 & 0 \cr 1 & 0 & 0 & 0 }
U^*_{(3/2)} \pmatrix{ 0 & 0 & 0 & 1 \cr 0 & 0 & -1 & 0 \cr
                      0 & 1 & 0 & 0 \cr -1 & 0 & 0 & 0 } ~.
\eeq
This can be seen by noting that, in terms of $u_{11}$ and $u_{12}$, the
elements in the first row of the SU(2) matrix in the fundamental
representation we have
\beq
U_{(3/2)} = \pmatrix{ u_{11}^3 & \sqrt{3} u_{11}^2 u_{12} &
 \sqrt{3} u_{11} u_{12}^2 & u_{12}^3 \cr
 - \sqrt{3} u_{11}^2 u^*_{12} & u_{11} (|u_{11}|^2 - 2 |u_{12}|^2) &
 u_{12} (2 |u_{11}|^2 - |u_{12}|^2) & \sqrt{3} u^*_{11} u_{12}^2 \cr
 \sqrt{3} u_{11} (u^*_{12})^2 & - u^*_{12} (2 |u_{11}|^2 - |u_{12}|^2) &
 u^*_{11} (|u_{11}|^2 - 2 |u_{12}|^2) & \sqrt{3} (u^*_{11})^2 u_{12} \cr
 - (u^*_{12})^3 & \sqrt{3} u^*_{11} (u^*_{12})^2 &
 - \sqrt{3} u^*_{12} (u^*_{11})^2 & (u^*_{11})^3 } ~.
\eeq

\noi
Again it is instructive to see the expected pattern of chiral symmetry
breaking also from the possible fermion bilinears that can be formed
out of staggered fermions carrying different representations. To start,
we generalize the argument of ref.~\cite{Simon}
and rewrite the massless staggered fermion
action for arbitrary real and pseudo-real representations, {\it i.e.,}
taking
\beq
U^*_\mu = S^\dagger U_\mu S \qquad {\rm with} ~ S S^* = \pm 1 ~.
\eeq
Using $\eta_\mu(x\pm\mu) = \eta_\mu(x)$ we can write
\beq
S_{st} = \frac{1}{2} \sum_{x=even,\mu} \eta_\mu(x) [ A(x) + B(x) ] ~,
\eeq
with
\beqn
A(x) &=& \bar \chi_e(x) U_\mu(x) \chi_o(x+\mu) -
 \bar \chi_e(x) U^\dagger_\mu(x-\mu) \chi_o(x-\mu) \nonumber \\
B(x) &=& \bar \chi_o(x-\mu) U_\mu(x-\mu) \chi_e(x) -
 \bar \chi_o(x+\mu) U^\dagger_\mu(x) \chi_e(x) ~.
\eeqn
Using the Grassmann nature of the $\chi$'s we can also express $B(x)$ as
\beqn
B(x) &=& -\chi_e^T(x) \left(U^\dagger_\mu(x-\mu)\right)^* \bar \chi_o^T(x-\mu)
 + \chi_e^T(x) U^*_\mu(x) \bar \chi_o^T(x+\mu) \nonumber \\
&=& \chi_e^T(x) S^\dagger U_\mu(x) S \bar \chi_o^T(x+\mu) -
 \chi_e^T(x) S^\dagger U^\dagger_\mu(x-\mu) S \chi_o^T(x-\mu) ~.
\eeqn
In the second line we have made use of the fact that the representation
is either real or pseudo-real. Introducing (the signs are for later
convenience, the upper one for real and the lower one for pseudo-real
representations)
\beq
\bar X_e = \pmatrix{ \bar \chi_e, & \pm \chi_e^T S^\dagger } ~, \qquad
X_o = \pmatrix{ \chi_o \cr \pm S \bar \chi_o^T }
\eeq
the massless staggered fermion action for real or pseudo-real representations
can be written as
\beq
S_{st} = \frac{1}{2} \sum_{x=even,\mu} \eta_\mu(x) \left[ \bar X_e(x)
U_\mu(x) X_o(x+\mu) - \bar X_e(x) U^\dagger_\mu(x-\mu) X_o(x-\mu)
\right] ~.
\eeq
For $N$ staggered fermions, we see that the usual $U(N)_e \times U(N)_o$
symmetry
\beqn
\chi_e \longmapsto W_e \chi_e &,&  ~~\bar \chi_o \longmapsto \bar \chi_o
 W_e^\dagger \qquad W_e \in U(N) \nonumber \\
\chi_o \longmapsto W_o \chi_o &,& ~~ \bar \chi_e \longmapsto \bar \chi_e
 W_o^\dagger \qquad W_o \in U(N)
\eeqn
is enlarged to $U(2N)$:
\beq
X_o \longmapsto W X_o ~,~~~ \bar X_e \longmapsto \bar X_e
 W^\dagger \qquad W \in U(2N) ~.
\eeq

\noi
Next we recast the condensate in terms of $X_o$ and $\bar X_e$:
\beqn
\bar \chi \chi &=& \frac{1}{4} \left[ \bar \chi_e \chi_e -\chi_e^T
 \bar \chi_e^T + \bar \chi_o \chi_o - \chi_o^T \bar \chi_o^T \right]
\nonumber \\
&=& \frac{1}{4} \left[ \bar X_e \pmatrix{ 0 & \unit \cr \mp \unit & 0 }
 S \bar X_e^T \mp X_o^T \pmatrix{ 0 & \unit \cr \mp \unit & 0 }
 S^\dagger X_o \right] ~,
\eeqn
where upper and lower signs are for real and pseudo-real representations,
respectively, and $\unit$ is the $N \times N$ unit matrix. The residual
symmetry left unbroken is that which leaves invariant the anti-symmetric
and symmetric $2N \times 2N$ form. Thus, for real representations the
chiral symmetry breaking pattern of staggered fermions is
\beq
U(2N) ~\to~ Sp(2N) ~,
\eeq
while for pseudo-real representations it is
\beq
U(2N) ~\to~ O(2N) ~.
\eeq
This is exactly the opposite of the continuum theory.
In table ~\ref{chisb} we summarize the results for both continuum 
and staggered fermions. 

\begin{table}
\caption{Chiral symmetry breaking patterns for continuum and staggered
fermions.}
\label{chisb}
\begin{center}
\begin{tabular}{|c|c|c|c|}
\hline
Rep $r$ & Fermions & Coset & RMT ens   \\
\hline
pseudo-real & Continuum & SU$(2N_f)/$Sp($2N_f$) & chOE \\
complex     & Continuum & SU($N_f$)             & chUE \\
real        & Continuum & SU$(2N_f)/$SO($2N_f$) & chSE \\
pseudo-real & Staggered & U$(2N)/$SO($2N$)      & chSE \\
complex     & Staggered & U($N$)                & chUE \\
real        & Staggered & U$(2N)/$Sp($2N$)      & chOE \\
\hline
\end{tabular}
\end{center}
\vskip 5mm
\end{table}

\noi
It is quite troubling that the patterns of spontaneous chiral symmetry
breaking for staggered fermions in all pseudo-real and real representations
are entirely different from those of continuum fermions. This is not
a minor issue, but concerns the whole spectrum of (pseudo-)Goldstone
bosons away from the continuum: at finite lattice spacing it is 
completely wrong. How can the correct
Goldstone spectrum possibly be recovered in the continuum limit? Let us
first do a simple counting of massless degrees of freedom.
In the case of real representation staggered
fermions the coset of chiral symmetry
breaking is U(2$N$)/Sp(2$N$). The dimension of this coset is $N(2N-1)$, and
there are as many Goldstone bosons away from the continuum. For pseudo-real
staggered fermions the coset is U(2$N$)/O(2$N$), the dimension of which
equals $N(2N+1)$. There are thus {\em more} Goldstone bosons in this case.
As the continuum limit is reached, these two cosets
should get interchanged. While it is fairly easy to imagine how more
massless states can appear in the continuum limit (away from the continuum
these states could have order-$a$ artifacts that disappear in the continuum
limit), it is difficult to imagine how strictly massless
modes at any finite lattice spacing can become {\em massive} in the continuum
limit. We find that the only resolution to this paradox is to assume
that the interchange of symmetry breaking classes occurs simultaneously
with the well-known enhancement of flavor symmetry: each staggered flavor
becomes 4 flavors of continuum fermions. In this way the number of strictly
massless states at finite lattice spacing will always be smaller than
that expected in the continuum\footnote{The needed inequality based on
a quadrupling of flavor degrees of freedom in the continuum limit
is $N(2N+1) < 4N(8N-1)-1$, which is satisfied for all positive integers
$N$.}, and all that happens is that nearly
massless states become massless in the continuum limit. Of course we have no
proof of this, and it is clearly also intimately tied up also with the anomaly
and the way the proper
flavor singlet remains massive in the continuum limit. In any case it
shows that we are extremely lucky that QCD is a theory of complex
representation fermions, for which the issue of wrong Goldstone manifolds
for staggered fermions is not relevant.

\section{Numerical Results}

\noi
Let us now turn to the results of our numerical simulations. Some details
about the ensembles are listed in Table ~\ref{tab:ens}. All ensembles are
pure gauge, generated with a mixture of overrelaxation and heatbath sweeps,
both working on the various SU(2) subgroups of the gauge group. The
ensembles were generated in the fundamental representation of the gauge group
with the standard Wilson action. For each of the three Random Matrix Theory
ensemble, and hence chiral symmetry breaking patterns, we have chosen one
example with a larger representation of the gauge group than the fundamental
one.

\begin{table}
\caption{Details of the gauge field ensembles, including gauge group,
$\beta$ value, volume, number of configurations analyzed, representation
considered (for gauge group SU(2) we label representations by isospin
$j$), predicted Random Matrix Theory ensemble and value of the
condensate $\Sigma$ obtained as explained in the text.}
\label{tab:ens}
\begin{center}
\begin{tabular}{|l|c|c|r|c|c|l|}
\hline
Group & rep & RMT ens & $\beta$ & $V$ & $N_{cfg}$ & $\Sigma$ \\
\hline
SU(2) & 3/2 & chSE & 2.2 & $4^4$ & 4623 & 1.234(60) \\
SU(3) & 6   & chUE & 5.1 & $4^4$ & 8000 & 3.569(18) \\
SU(3) & 6   & chUE & 5.1 & $6^4$ & 4777 & 3.609(29) \\
SU(4) & 6   & chOE &10.3 & $4^4$ & 9000 & 2.361(22) \\
SU(4) & 6   & chOE &10.3 & $6^4$ & 5840 & 2.468(24) \\
SU(4) & 4   & chUE &10.3 & $4^4$ & 9000 & 1.046(9)  \\
\hline
\end{tabular}
\end{center}
\vskip 5mm
\end{table}

\noi
We computed the low-lying part of the spectrum of the staggered Dirac
operator in the desired representation as described in Ref.~\cite{DHNR}.
If chiral symmetry breaking occurs, the rescaled eigenvalues $\zeta = V
\Sigma \lambda$ should have the microscopic spectral density
eq.~(\ref{rhodef}) corresponding to the appropriate chiral Random Matrix
Theory ensemble. Similarly, the rescaled lowest eigenvalue should be
distributed according to the distribution $P_{\rm min}(\zeta)$ of the
same chiral Random Matrix Theory ensemble. Of course, to make the
comparison, one needs to know the value $\Sigma$ of the chiral
condensate. If the distribution of the lowest eigenvalue agrees, in shape,
with the predicted distribution, then $\Sigma$ can be obtained from the
rescaling that gives the best quantitative agreement between measured and
theoretically predicted distribution. Alternatively, and somewhat simpler,
one can obtain $\Sigma$ from the average of the lowest eigenvalue,
$\langle \lambda_{\rm min} \rangle$ and the theoretical mean value
$\overline{\zeta}$
\beq
\overline{\zeta} = \int_0^\infty \zeta P_{\rm min}(\zeta) d\zeta
\label{mean}
\eeq
as $\Sigma = \overline{\zeta} / (V \langle \lambda_{\rm min} \rangle)$.
The values obtained in this way are also given in Table~\ref{tab:ens}.
Some of the very earliest studies of chiral symmetry breaking and
chiral symmetry restoration at finite temperature \cite{Kogut} also
studied some of these exotic quark representations, but without aiming
at answering the present questions.

\noi
One important issue regarding staggered fermions is the comparison
with analytical results in sectors of fixed topological charge $\nu$.
For staggered fermions it is normally argued that at present-day
gauge couplings these lattice fermions are insensitive to quantum
gauge field configurations of non-trivial topological charge. Indeed,
even if one restricts oneself to gauge field sectors that clearly carry
non-zero topological charge, one finds that staggered fermions behave
as if the charge $\nu$ were zero \cite{DHNR}. Consistent with this
is the observation that the fermion zero modes that by the index theorem
are present for continuum fermions in non-trivial sectors of topological
charge appear completely mixed up with the other small Dirac operator
eigenvalues \cite{DHNR}. Actually, there is an alternative way of 
understanding these results \cite{D01}. Note that we found that
all cosets of spontaneous chiral symmetry breaking for staggered fermions 
differ by one additional U(1) factor from the cosets of the continuum. 
This additional U(1) factor plays a crucial role in determining the
distributions of the smallest Dirac operator eigenvalues: to leading
order in the finite-volume chiral Lagrangian expansion the integration over
a coset enlarged by an additional U(1) factor is simply {\em equivalent} to
a projection onto the $\nu=0$ sector of the theory without this U(1)
factor. This is easily illustrated. Consider, for example, the universality
class of $\beta=2$, for which the smallest Dirac operator eigenvalues
of continuum fermions are determined by the zero-momentum modes of the 
following effective partition function \cite{LS}:
\beq
{\cal Z} ~=~  \int_{SU(N_f)}\! dU~ \exp\left[mV\Sigma {\mbox{\rm Re Tr}}
\left(e^{i\theta/N_f}{\cal M}U^{\dagger}\right)\right] ~,
\eeq
where $\theta$ is the vacuum angle, and ${\cal M}$ is the quark mass matrix. 
Projection onto a sector of topological charge $\nu=0$ yields
\beq
{\cal Z}_0 ~=~  \int_{U(N_f)}\! dU~ \exp\left[mV\Sigma {\mbox{\rm Re Tr}}
\left({\cal M}U^{\dagger}\right)\right] ~,
\eeq
which is completely equivalent to the partition function based on the
coset U$(N_f)\times$U($N_f)/$U($N_f$) at vanishing vacuum angle. We learn
that the additional U(1) factor of staggered fermions is entirely
equivalent to a projection down on the $\nu=0$ sector of the 
theory without this U(1) factor, at $\theta=0$. 
The two other symmetry breaking
patterns of staggered fermions, having also such additional U(1) factors
in their cosets (see table ~\ref{chisb} for a summary), 
are completely similar in this respect. Therefore,
independently of the more empirical observations of ref.
\cite{DHNR}, we simply {\em should} compare with the predictions of
the $\nu=0$ sectors of the theories without these additional U(1) symmetries. 
It should also be remarked
that this equivalence between the predictions for staggered fermions
and those of theories without the U(1) symmetries 
for $\nu=0$ is only valid to leading
order in the finite-volume chiral expansion \cite{D01}. Already at
the next-to-leading order differences show up, but we do not have
sufficient statistics to probe these small deviations here. 
 
\begin{figure}
\begin{tabular}{c c}
\includegraphics[scale=0.5]{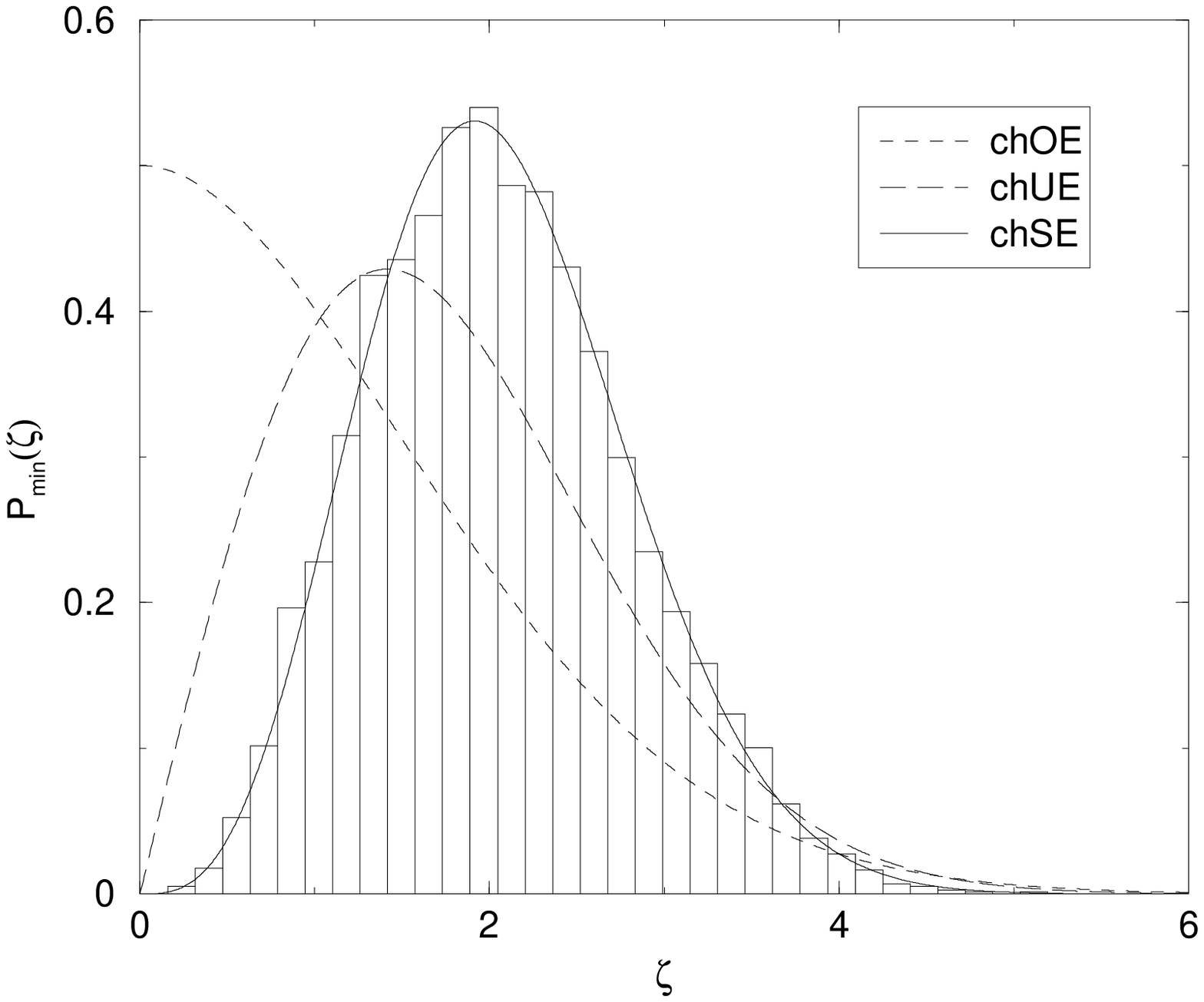}
&
\includegraphics[scale=0.5]{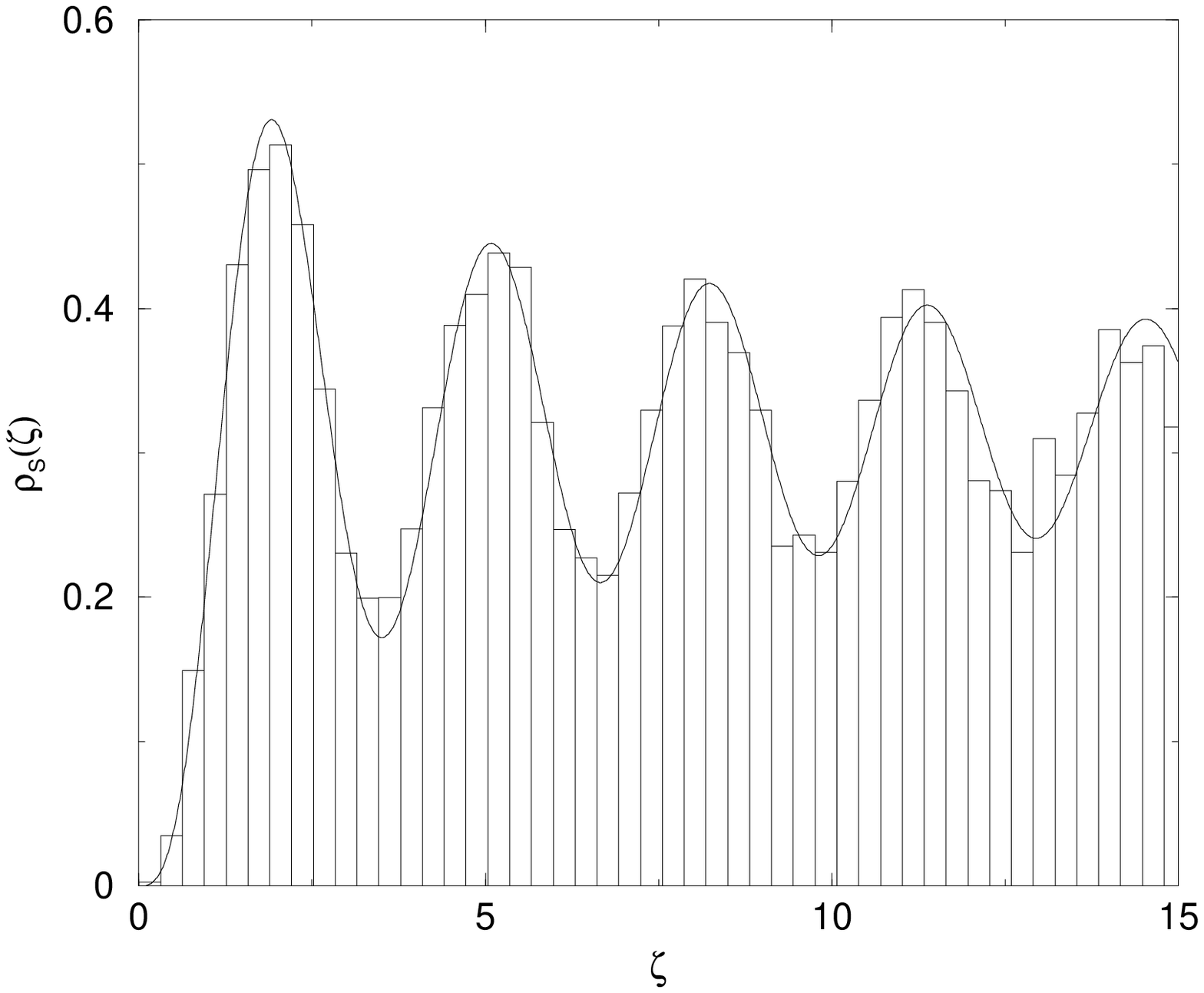}
\end{tabular}
\caption{The left panel shows the microscopic distribution of the first
eigenvalue of the SU(2) ensemble in the $j=3/2$ representation, which is
pseudo-real.
The analytic predictions for the three ensembles chOE, chUE and
chSE are included. The data clearly follow the curve for the (chiral)
symplectic ensemble, as expected. In the right panel the beginning of the
full microscopic spectral density of the same ensemble is displayed. 
The space-time volume used is $V=4^4$.}
\label{fig:su2_4}
\end{figure}

\begin{figure}
\begin{tabular}{c c}
\includegraphics[scale=0.5]{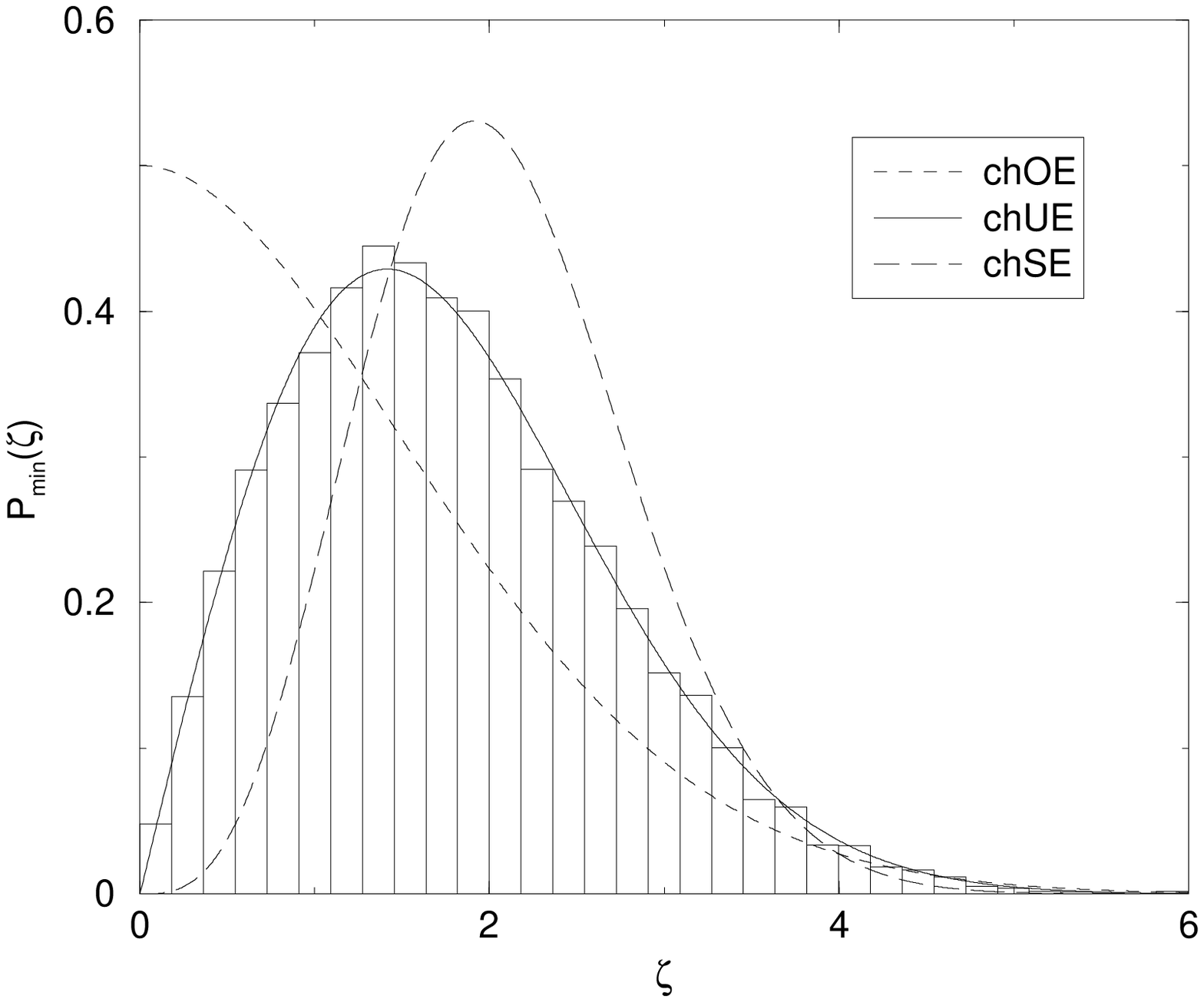}
&
\includegraphics[scale=0.5]{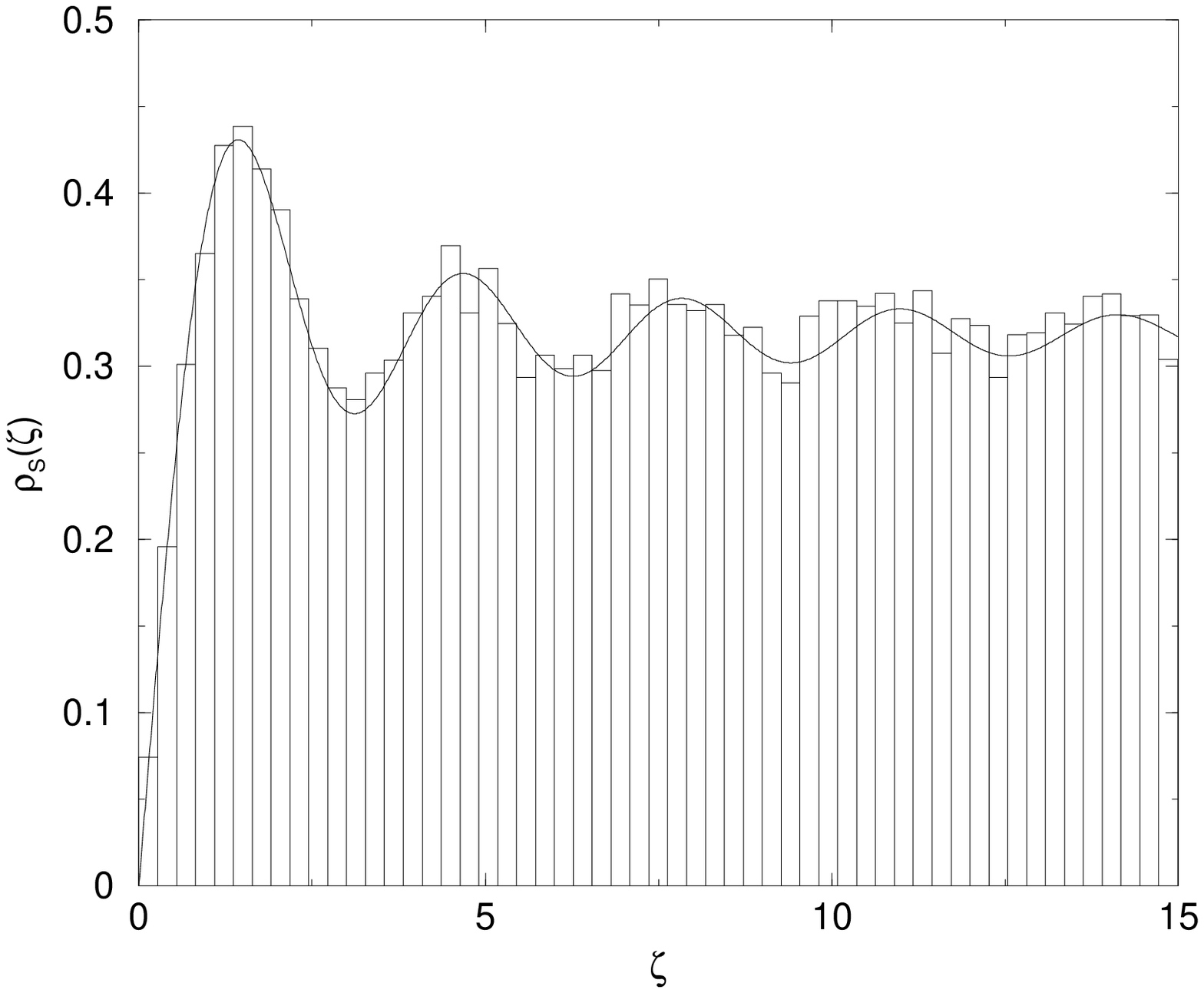}
\end{tabular}
\caption{The microscopic distribution of the first eigenvalue (left) and
the spectral density (right) for the SU(3) ensemble in the 6-representation
(which is complex),
from a $4^4$ lattice. The data clearly follow the prediction for the (chiral)
unitary ensemble.}
\label{fig:su3_4}
\end{figure}

\begin{figure}
\begin{tabular}{c c}
\includegraphics[scale=0.5]{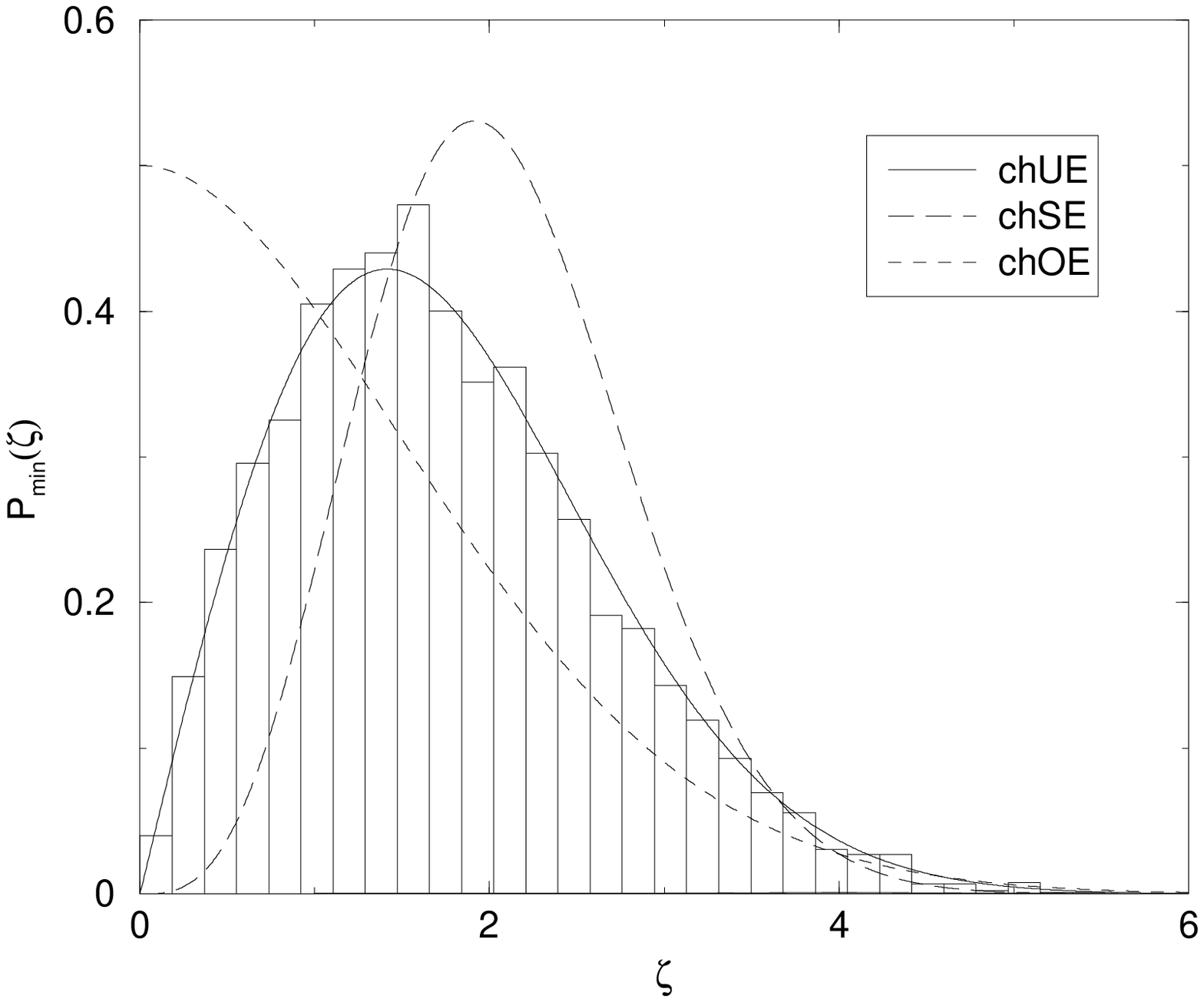}
&
\includegraphics[scale=0.5]{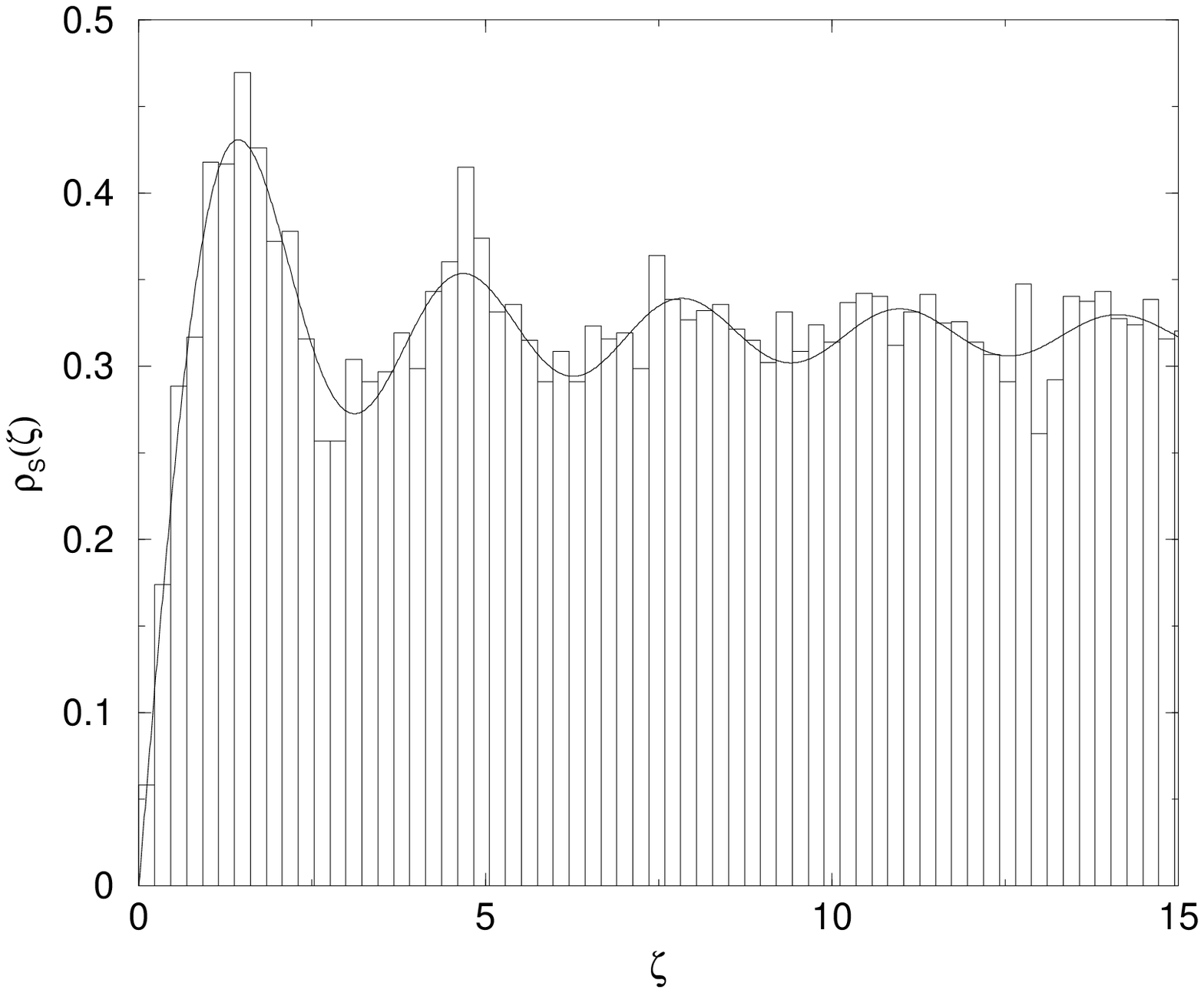}
\end{tabular}
\caption{Same as Fig.~\ref{fig:su3_4} but for a $6^4$ lattice. The fact
that data are fit with the same parameter $\Sigma$ shows that the
eigenvalues indeed scale with the volume as required for spontaneous
chiral symmetry breaking.}
\label{fig:su3_6}
\end{figure}

\begin{figure}
\begin{tabular}{c c}
\includegraphics[scale=0.5]{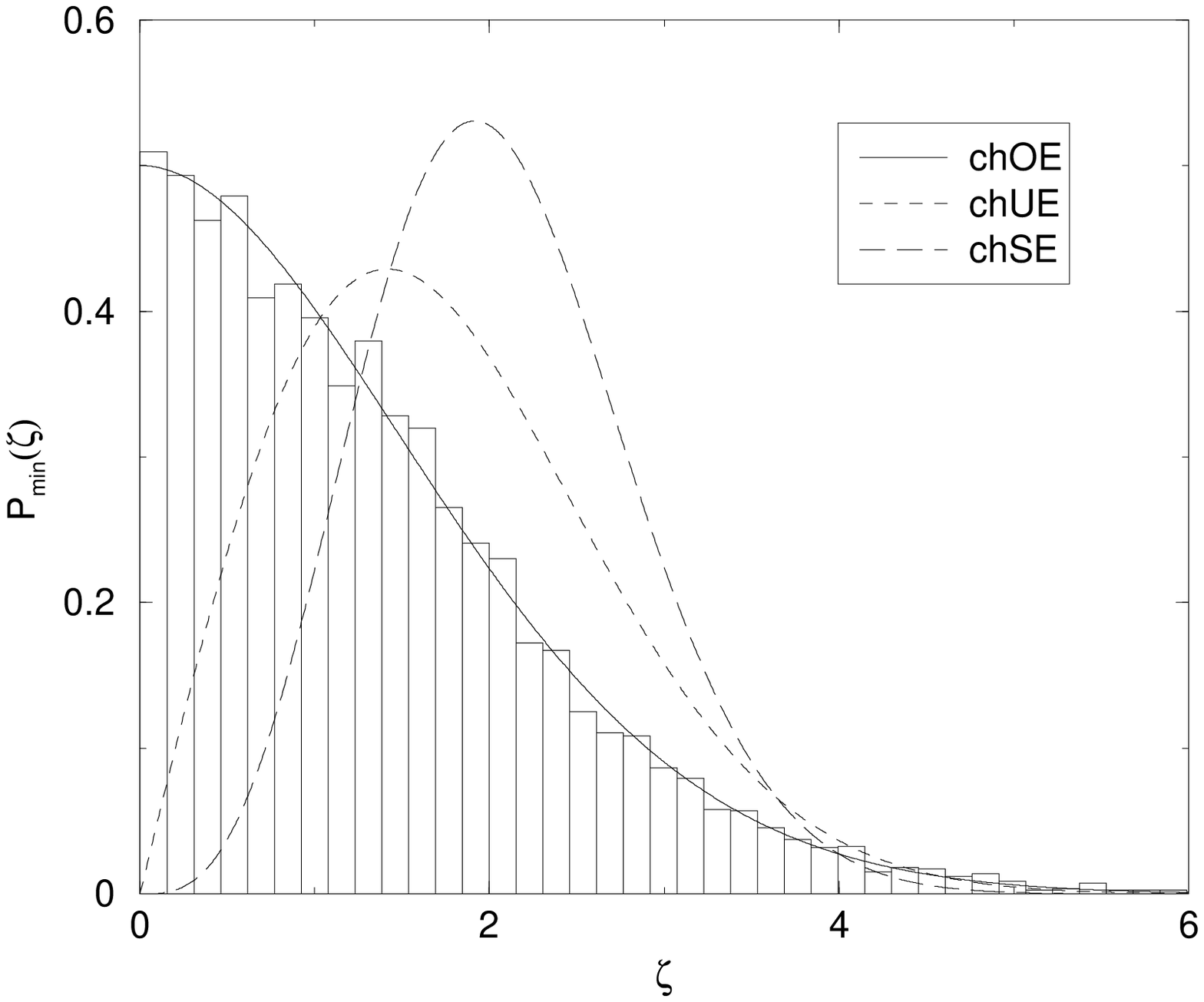}
&
\includegraphics[scale=0.5]{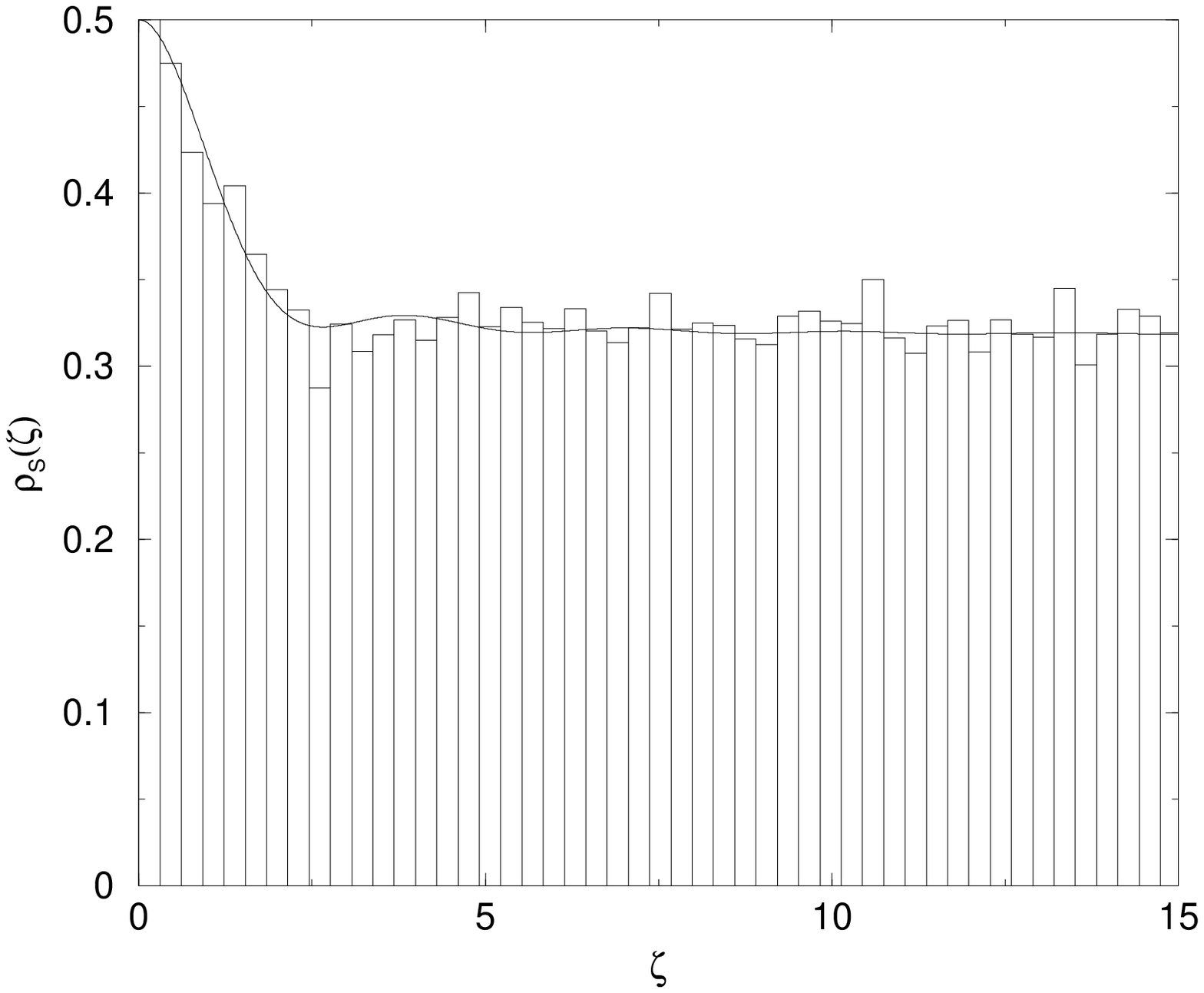}
\end{tabular}
\caption{The microscopic distribution of the first eigenvalue (left) and
the spectral density (right) for the SU(4) ensemble in the anti-symmetric
6-representation (which is real), from a $4^4$ lattice. 
The data clearly follow the
prediction for the (chiral) othogonal ensemble.}
\label{fig:su4_4}
\end{figure}

\begin{figure}
\begin{tabular}{c c}
\includegraphics[scale=0.5]{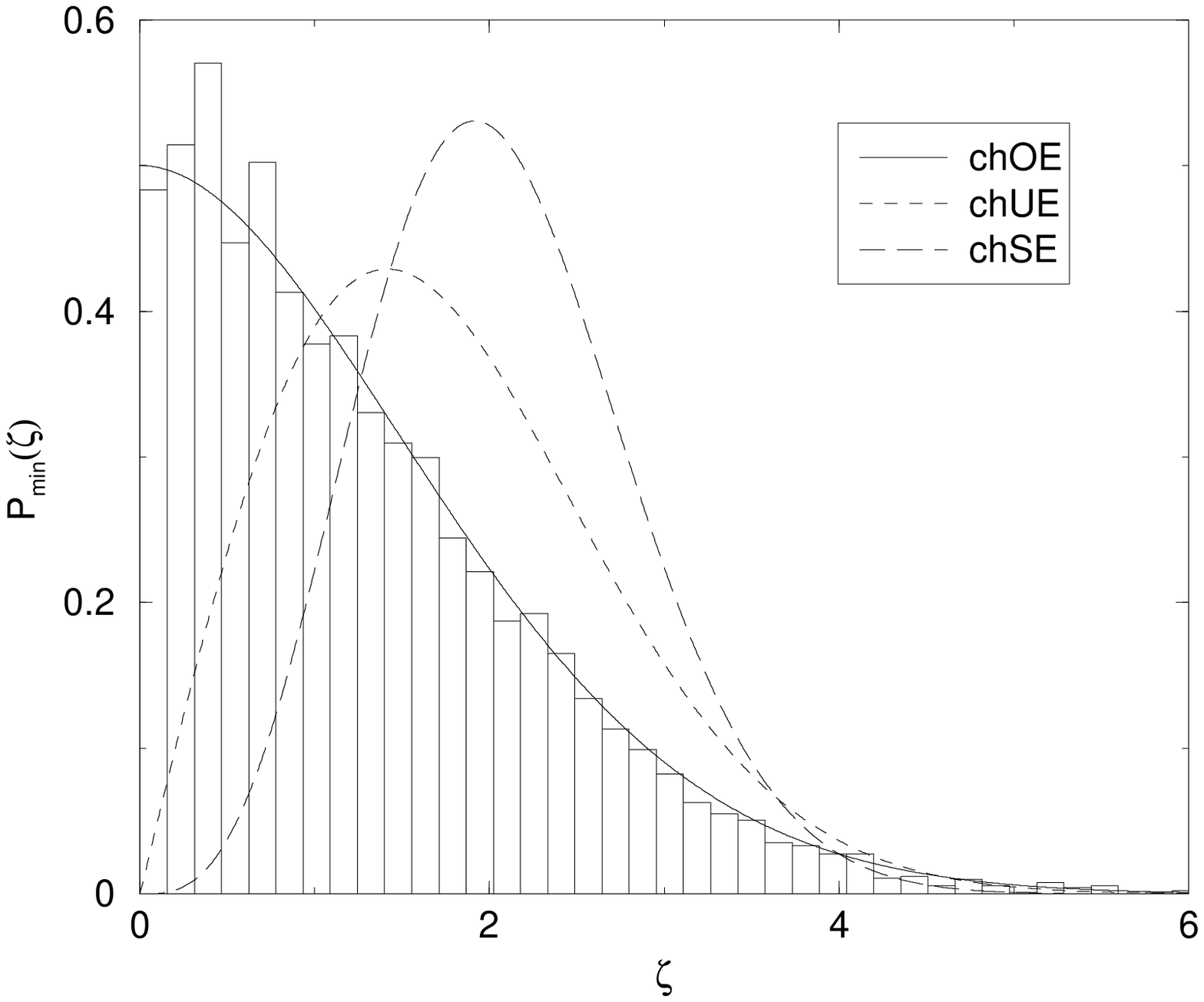}
&
\includegraphics[scale=0.5]{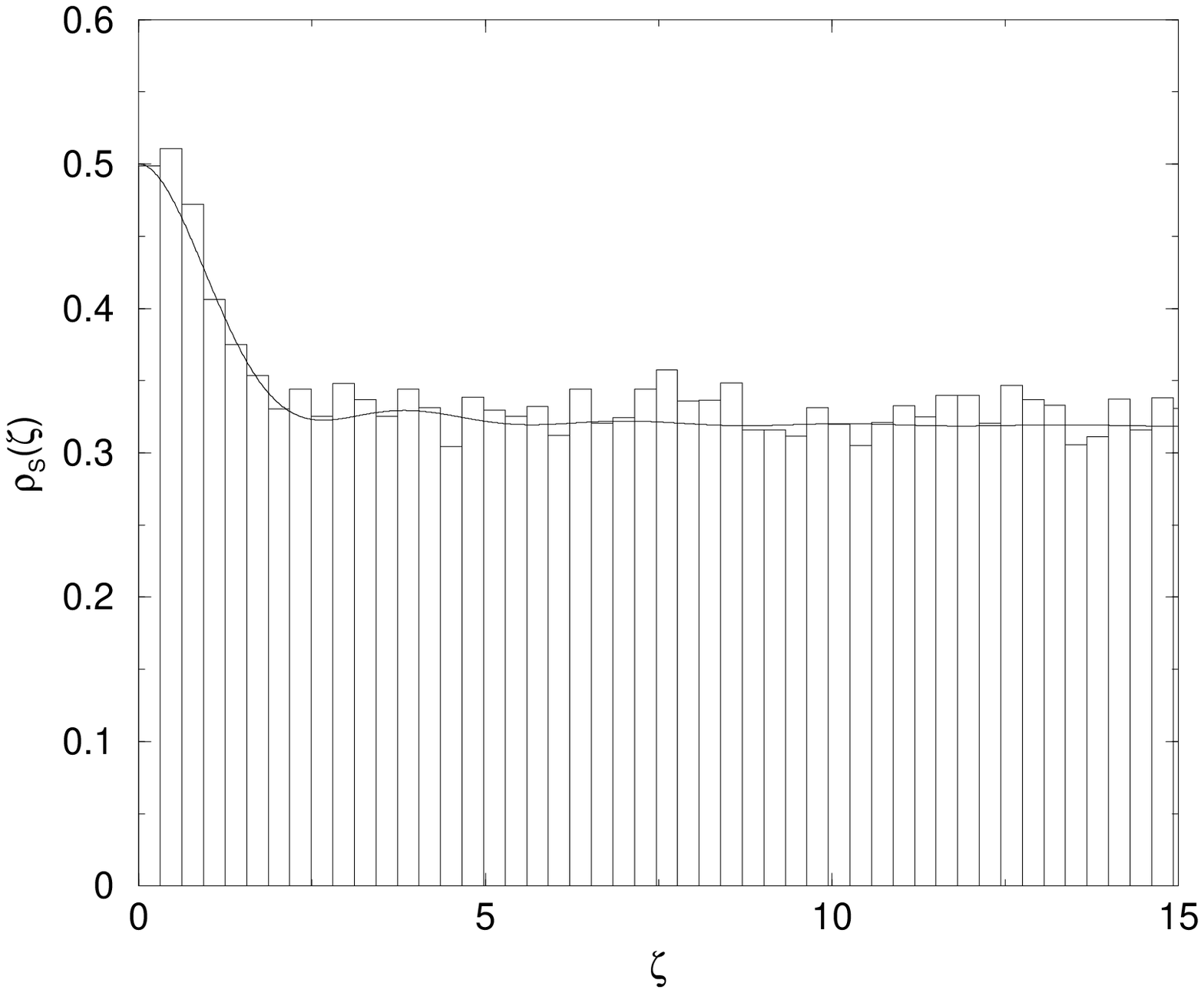}
\end{tabular}
\caption{Same as Fig.~\ref{fig:su4_4} but for a $6^4$ lattice.}
\label{fig:su4_6}
\end{figure}

\noi
The comparison of the measured and predicted distributions (recalling
the remarks above) is shown in Figs.~\ref{fig:su2_4} to \ref{fig:su4_6}. 
For the smallest eigenvalue distributions we should thus compare
with the $\nu=0$ predictions \cite{NDW,F}
\begin{eqnarray}
P_{{\mbox{\rm min}}}(\zeta) &=& \frac{\zeta+2}{4}e^{-\zeta/2-\zeta^2/8}~~,
~~~~~~~ \beta = 1 \cr
P_{{\mbox{\rm min}}}(\zeta) &=& \frac{\zeta}{2}e^{-\zeta^2/4}~~,~~~~~~
\beta=2 \cr
P_{{\mbox{\rm min}}}(\zeta) &=& \sqrt{\frac{\pi}{2}}\zeta^{3/2}I_{3/2}(\zeta)
e^{-\zeta^2/2}~~,~~~~~~~ \beta=4 ~.
\end{eqnarray}
These eigenvalue distributions yield the following mean values 
(see eq. (\ref{mean})) $\bar{\zeta}$: 1.311.. (for $\beta=1$), 1.772.. (for
$\beta=2$), and 2.066.. (for $\beta=4$). 

\noi
Similarly, for the microscopic spectral densities we should compare
with \cite{VZ,V,NF,NN,AK}
\begin{eqnarray}
\rho_s(\zeta) &=& \frac{\zeta}{2}\left(J_0(\zeta)^2+J_1(\zeta)^2\right)
-\frac{1}{2}J_0(\zeta)\left(\int_0^{\zeta}\!dt~J_0(t)-1\right)~~, ~~ \beta=1
\cr
\rho_s(\zeta) &=& \frac{\zeta}{2}\left(J_0(\zeta)^2+J_1(\zeta)^2\right)~~,
~~\beta=2 \cr
\rho_s(\zeta) &=& \zeta\left(J_0(2\zeta)^2+J_1(2\zeta)^2\right)
-\frac{1}{2}J_0(2\zeta)\int_0^{2\zeta}\!dt~J_0(t)~~, ~~ \beta=4
\end{eqnarray}
for the three different universality classes indicated.
In all the figures we see
beautiful agreement between the measured distributions and the theoretical
predictions. 

\noi
In two cases, for the 6-representations of SU(3) and SU(4), we considered
two different volumes, in order to also illustrate the finite-size scaling
that allows us to infer that indeed chiral symmetry is spontaneously broken.
One quantitative measure of this is that the extracted value for the 
parameter $\Sigma$ (which should be identified with the infinite-volume
chiral condensate)
should be independent of the volume. As can be seen in Table~\ref{tab:ens}
the agreement between the values obtained on the different lattice sizes
is very satisfactory. Altogether we have thus not only confirmed the 
theoretical prediction for
the patterns of chiral symmetry breaking, but also shown that chiral
symmetry breaking does indeed occur. The agreement is only possible
if the average spacing between eigenvalues near the origin goes like
$1/(V\rho(0))$ and therefore, in the infinite volume limit, a non-zero
density $\rho(0)$ is built up.

\begin{figure}
\begin{tabular}{c c}
\includegraphics[scale=0.5]{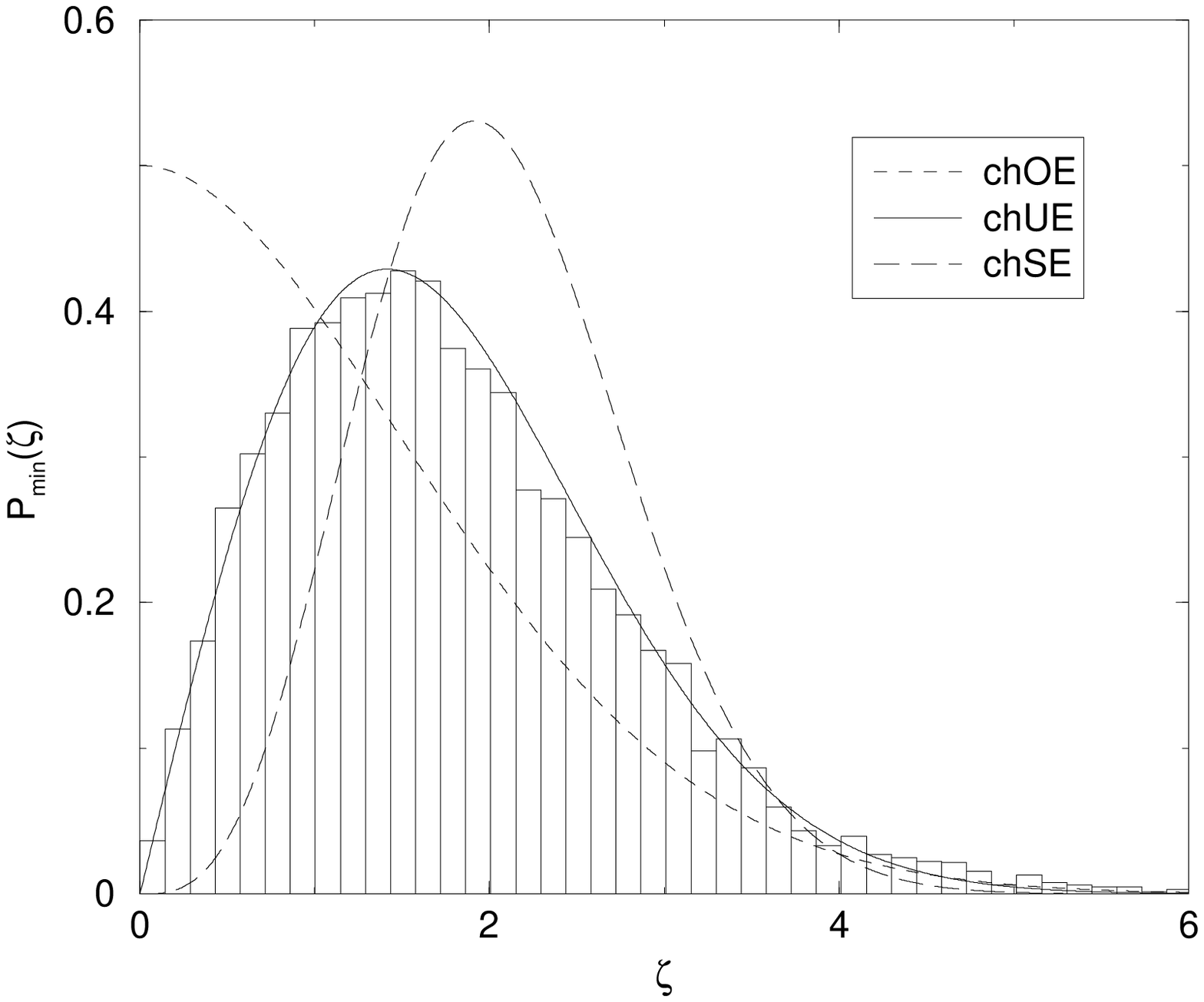}
&
\includegraphics[scale=0.5]{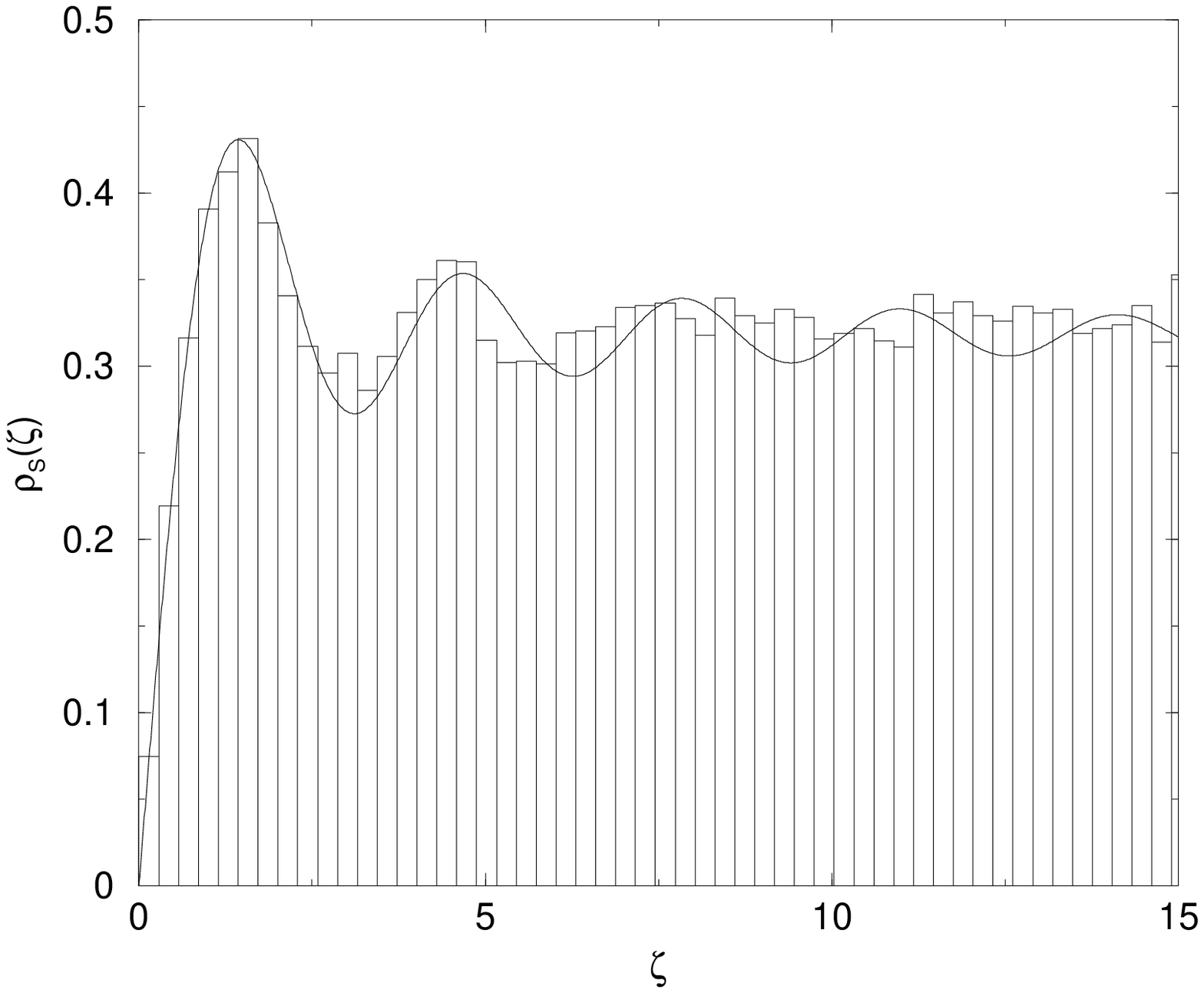}
\end{tabular}
\caption{The microscopic distribution of the first eigenvalue (left) and
the spectral density (right) for the SU(4) ensemble in the fundamental
representation (which is complex), 
from a $4^4$ lattice. The data clearly follow the
prediction for the (chiral) unitary ensemble.}
\label{fig:su4f_4}
\end{figure}

\noi
Since this is the first study of the staggered eigenvalue distributions for
gauge group SU(4) we decided to also look at the fundamental representation
for which the corresponding Random Matrix Theory ensemble should be chiral
unitary. As seen in Fig.~\ref{fig:su4f_4} the agreement between
measured distributions and the theoretical prediction is very nice here 
as well.

\section{Conclusions}

\noi
We have investigated the idea that there are only three distinct classes of
spontaneous chiral symmetry breaking of vectorlike gauge theories
in four dimensions. A particular issue concerns the lattice regularization
of the fermions. It has been known for quite a while that staggered
fermions in the fundamental representation of SU(2) have a spontaneous
symmetry breaking pattern that falls in a wrong category as compared
with continuum fermions, and it has likewise been known that staggered
fermions in the adjoint representation of any SU(N) gauge group have
a pattern of symmetry breaking
that also falls in a wrong category compared with
continuum fermions. Here we have generalized this argument to any
irreducible representation $r$ of any simple gauge group ${\cal G}$.
It turns out that for all real and for
all pseudo-real representations the symmetry breaking classes for
staggered fermions are simply swapped as compared with continuum fermions.
We have given some suggestions as to how the correct symmetry breaking
classes may be recovered in the continuum limit.

\noi
We hope to have given convincing evidence that an extremely efficient tool
for determining whether spontaneous chiral symmetry breaking occurs,
and for simultaneously finding the universality class
to which it belongs, is provided by the distributions of the smallest
Dirac operator eigenvalues. As a by-product one obtains not only the
answer to these two questions, but also an accurate determination of
$\Sigma$, the infinite-volume chiral condensate. The method here is
essentially a finite-size scaling analysis, where in the case of
Dirac operator eigenvalues one actually knows the exact analytical
form of the scaling function. By going one step further, and looking
carefully at the corrections to finite-volume scaling, one can even
extract one more parameter, the pion decay constant $F$, from just
these eigenvalue distributions \cite{D01}.

\noi
To illustrate the technique, we have performed Monte Carlo simulations
of exotic representations of SU(2), SU(3) and SU(4) gauge groups
with staggered fermions. In all cases we have found complete agreement with
the expected patterns of chiral symmetry breaking. This, taken together
with the four other known cases in the literature \cite{BBetal,Lang}, puts
the conjecture of symmetry breaking classes on a quite solid footing.

\noi
{\sc Acknowledgement:}\\
The work of PHD. was supported in part by EU TMR grant no. ERBFMRXCT97-0122,
and the work of UMH by DOE contract DE-FG02-97ER41022.
The work of BS was supported in part by the U.S. Department of Energy under
Contract No. DF-FC02-94ER40818.  He thanks the Center for Theoretical
Physics at MIT for their hospitality during his sabbatical leave from Tel
Aviv University. PHD, UMH and BS also acknowledge the financial support of 
NATO Collaborative Linkage Grant PST.CLG.977702.

\end{document}